\documentclass{aa}  

\usepackage{graphicx}
\usepackage{txfonts}
\usepackage{natbib}   

\usepackage{hyperref}
\begin{document} 


  \title{Parker Solar Probe detects solar radio bursts related\\ with a behind--the--limb active region}
        \titlerunning{Parker Solar Probe detects solar radio bursts related with a behind--the--limb AR}


   \author{Aleksander~A.~Stanislavsky\inst{1,2}
          \and
          Igor~N.~Bubnov\inst{1,3}
          \and
          Artem~A.~Koval\inst{4}
          \and
          Serge~N.~Yerin\inst{1,3}
          }

   \institute{Institute of Radio Astronomy, National Academy of Sciences of Ukraine, Mystetstv St., 4, 61002 Kharkiv, Ukraine\\
              \email{a.a.stanislavsky@rian.kharkov.ua}
         \and
             Faculty of Pure and Applied Mathematics, Hugo Steinhaus Center, Wroc{\l}aw University of Science and Technology,\\ Wyb.
Wyspia\'{n}skiego 27, 50-370 Wroc{\l}aw, Poland
          \and
             V.N. Karazin Kharkiv National University, Svobody Sq., 4, 61022 Kharkiv, Ukraine
          \and
             Astronomical Institute, Czech Academy of Sciences, 251 65 Ond\v{r}ejov, Czech Republic
             }

   \date{Received ; Accepted }

 
  \abstract
   {The interpretation of solar radio bursts observed by Parker Solar Probe (PSP) in the encounter phase plays a key role in understanding intrinsic properties of the emission mechanism in the solar corona. Lower time--frequency resolution of the PSP receiver can be overcome by simultaneous ground--based observations using more advanced antennas and receivers.}
   {In this paper we present such observations for which the active active region 12765, begetter of type III, J, and U solar bursts, was within sight of ground--based instruments and behind the solar limb of the PSP spacecraft.}
   {We used a subarray of the Giant Ukrainian Radio Telescope (GURT) to get the spectral properties of radio bursts at the frequency range of 8--80 MHz, as well as the PSP radio instruments with a bandwidth of 10.5 kHz -- 19.2 MHz, during solar observations on June 5, 2020.}
   {We  directly detected the radio events initiated by the active region behind the solar limb of the PSP spacecraft, using special conditions in the solar corona, due to the absence of active regions from the PSP side. Following the generation mechanism of solar radio emission, we refined the density model for the solar corona above the active region 12765 responsible for the radio bursts. Based on the PSP spacecraft position near the Sun and delays of radio waves between space-- and ground--based records, we found the corresponding radio responses on the PSP spectrogram.}
   {The absence of sunspots from the PSP side contributes to the propagation of radio waves from a dense loop of the Sun to quiet regions with low densities, through which PSP instruments can detect the radiation.}


   \maketitle

\section{Introduction}
The largest population of solar bursts, type III radio bursts, is a signature of electron beams propagating outward from lower layers of corona to interplanetary space along open magnetic field lines. They have been observed since the 1950s \citep{wild50} and are the most studied phenomenon among other types of solar radio emission from observational and theoretical points of view \citep{Reid14}. The main generation mechanism of type III radio bursts is that beam electrons with velocities of about 0.3$c$ ($c$ is the velocity of light) can induce Langmuir waves along their motion path, and the waves, scattered on ions, can be transformed into radio emission \citep[{e.g.,}][]{zhel70}. Observations of the solar radio events are provided in a very wide frequency range, from 1--3 GHz down to 10 kHz, by various space-- and ground--based instruments. 

Recently, a new spacecraft mission, named Parker Solar Probe (PSP), was sent to the Sun. Moving along elliptic orbits, the observatory approaches the Sun, then it moves away from it. This allows for radio emission to be received fairly close to the Sun. The low--frequency cutoffs for type III bursts were investigated with PSP during its encounters 1--5. The characteristic feature is caused by intrinsic properties of the type III burst emission mechanism as well as by propagation effects. The cutoff frequencies from PSP observations were higher than ones measured by Wind/WAVES and Ulysses missions \citep{Ma21}. Several type III bursts observed at the beginning of April 2019 were studied by \cite{Krupar20} in the frequency range of 10 kHz -- 19 MHz. A more comprehensive analysis of type III and type IIIb bursts at the same time by using both PSP and ground-based radio telescopes was presented by \citet{Melnik21}, showing that ground--based radio observations help to better understand results of PSP. In particular, according to them, PSP unequivocally observed the harmonic, but not the fundamental, component. The aim of this paper is to use the radio data recorded by PSP within 0.15 au to study solar low--frequency radio emission. Unlike previous studies, we have detected the U+J+III association that occurred on June 5, 2020. Its study by PSP will be considered in the present work.

This paper is organized as follows. Briefly starting with the PSP spacecraft and instruments installed on it, we describe radio observations carried out with the Giant Ukrainian Radio Telescope (GURT) (Kharkiv) on June 5, 2020. With the help of the ground--based measurements, our analysis of this solar event leads to the exact identification of solar bursts presented in it. This allows us to specify the density model above the active region (AR) responsible for the solar bursts. Knowing the PSP position, we show that the source emitting the bursts was initiated by electron beams from the AR behind the limb of the spacecraft. Finally, we discuss possible reasons of this detection and summarize our finding.

\section{Observations and facilities}

\subsection{PSP spacecraft and its instruments}
PSP is a new NASA spacecraft that launched on August 12, 2018. Its purpose is to fly to a closer distant to the Sun than any previous spacecraft \citep{Fox16}. In the maximum perihelion, PSP's elliptic orbit will be within 9.86 solar radii from the center of the Sun. This case will occur on December 24, 2024. Therefore, the spacecraft will be the first to measure, in situ, the solar corona properties in the birthplace of the solar wind. Direct measurements of electric and magnetic fields, radio waves, Poynting flux, absolute plasma density, as well as electron temperature are provided by the FIELDS instrument on PSP \citep{Bale16}. The FIELDS sensors include four whip antennas (called $V1$, $V2$, $V3$, and $V4$) mounted near the edge of the PSP heat shield. Each of them is 2 meters long and made of niobium alloy. The antennas are used with various measuring systems taking the raw signals and converting them into data for transmission back to Earth by radio communication. Radio observations are processed by the dual-channel receiver \citep{Pulupa17}. One Low Frequency Receiver (LFR) covers the bandwidth of 10.5 kHz -- 1.7 MHz, and another High Frequency Receiver (HFR) has the bandwidth of 1.3 MHz -- 19.2 MHz. Together they produce the data with the bandwidth of 10.5 kHz -- 19.2 MHz. When PSP is closer to the Sun than 0.25 au, all of its instruments work continuously at a high--rate recording mode. This is an encounter phase, whereas in a cruise phase, when the distance of PSP from the Sun is larger than 0.25 au, the data--recording rate becomes low. During the encounter phase, LFR and HFR record data at a cadence of one spectrum per $\sim$7 s. On the contrary, the data--recording cadence is noticeably less, one spectrum per $\sim$56 s. As of June 5, 2020, PSP was in perihelion at the distance of $\sim$0.15 au from the Sun. This case is just the encounter phase.

\subsection{Ground-based instruments}
Ground-based observational facilities included one array section of GURT, located near Kharkiv (Ukraine). It consists of 25 cross-dipoles, five being in each column and row. Dipole arms have north-south and east-west orientations. This antenna instrument permits radio emission to be received within the frequency range of 8--80 MHz \citep{Konov16}. Radio records were made with the help of an advanced digital receiver (ADR), which is a standard device of  GURT \citep{Zakharenko16}.  Its time resolution was about 100 ms in the frequency resolution of 38.147 kHz. To confirm our observational results, we also used other radio instruments: e--Callisto network stations, the Radio Solar Telescope Network (RSTN) observatory in San Vito of Italy, and the Nan\c{c}ay Decametric Array (NDA) in France. Their radio data are freely available for studies, but their time--frequency resolution is less than what GURT can offer. Unfortunately, we could not use the excellent Ukrainian T-shaped Radio telescope (UTR--2), operating in the frequency range of 9--33 MHz \citep{Konov16}, for observations.

\begin{figure*}[!h]
\begin{center}
   \includegraphics[width=14cm,keepaspectratio]{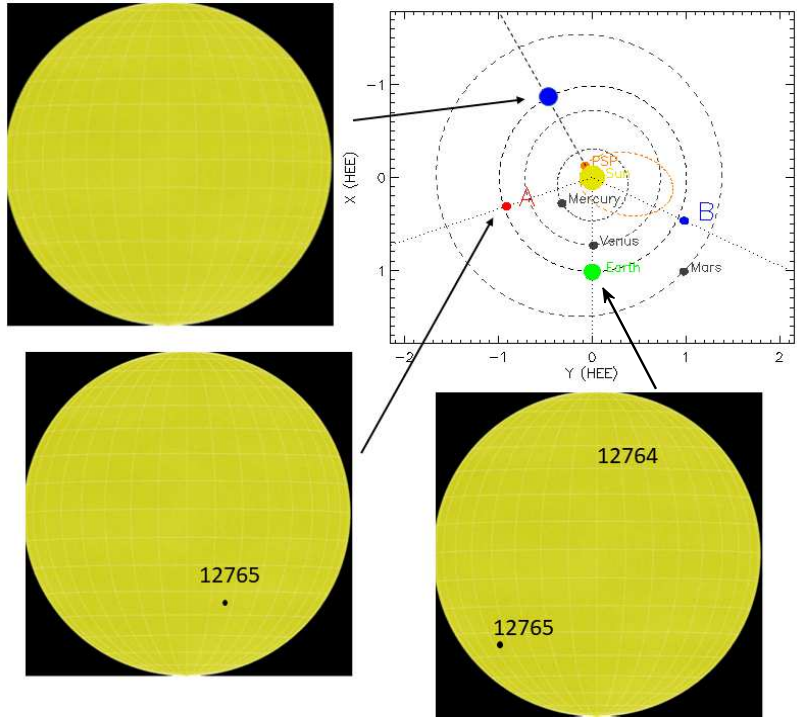}
\end{center}
\caption{Positions of PSP as well as STEREO A and B on June 5, 2020. We provide a solar photosphere view from three points on Earth's orbit as well as corresponding space-- and ground--based observers. Using the observations from the Helioseismic and Magnetic Imager (HMI) of the Solar Dynamics Observatory (SDO), the solar photosphere pictures were reproduced at three places as it would be seen along Earth's orbit.}
\label{fig1} 
\end{figure*}

\subsection{Solar events and observations}
In the summer of 2020, the Sun was quiet most of the time because of solar activity being minimal. Nevertheless, on May 29, 2020, the sunspot group produced the largest solar flare, starting from October 2017, when several X--class flares of September 2017 were observed. The flare on May 29, 2020, being of the M class, can be considered as the awakening of the Sun after a long inactivity. We would like to highlight that the smallest flares are a part of the A class, whereas the X class manifests the most intense events. The letters A, B, C, M, and X with decimal subclasses describe all the set of flares detected from the Sun. The M--class flare was a precursor of solar radio bursts in early June 2020. The activity was connected with a new active area, named NOAA AR 12765, which appeared on the limb from the eastern side of the Sun on June 3, 2020. The evolution of the active region started with a single spot (https://www.spaceweatherlive.com/en/solar-activity/region/12765), which became a small bipolar region on June 5, 2020, not being transformed into something more complex. On that day, the size of sunspots was the largest,  130~MH (in ``millionths of the visible solar hemisphere''). We note that the sunspot area was comparable to the entire surface area of the Earth having almost 170~MH. In the following days, the region remained bipolar, but started to decay after 10 June it became unipolar until it vanished completely, moving to the far side of the Sun as viewed from Earth. As the activity of solar regions correlates with their size, one can assume that a bipolar active region with a size of more than 110~MH will be able to generate high initial electron beam densities, resulting in radio emission of an U-burst with type III bursts. Solar observations on June 5, 2020 were also provided with space--based observatories: PSP; the Wind/WAVES instrument at the $L_1$ Lagrangian point; and STEREO--A (Solar Terrestrial Relations Observatory), orbiting around the Sun at a distance of 1 au ahead of Earth. They looked at the Sun from different points of view, with Earth's orbit and the PSP orbit near the Sun (Fig.~\ref{fig1}). For the PSP spacecraft, the AR 12765 was behind the solar limb ($\sim$106$^\circ$) with respect to the central meridian of the Sun. As for STEREO--A, Wind/WAVES, and ground--based observers, they could clearly see the active region on the solar photosphere. We note that the spacecraft STEREO--B (behind Earth) was out of contact.

\begin{figure*}[!h]
\begin{center}
         \includegraphics[width=14cm,keepaspectratio]{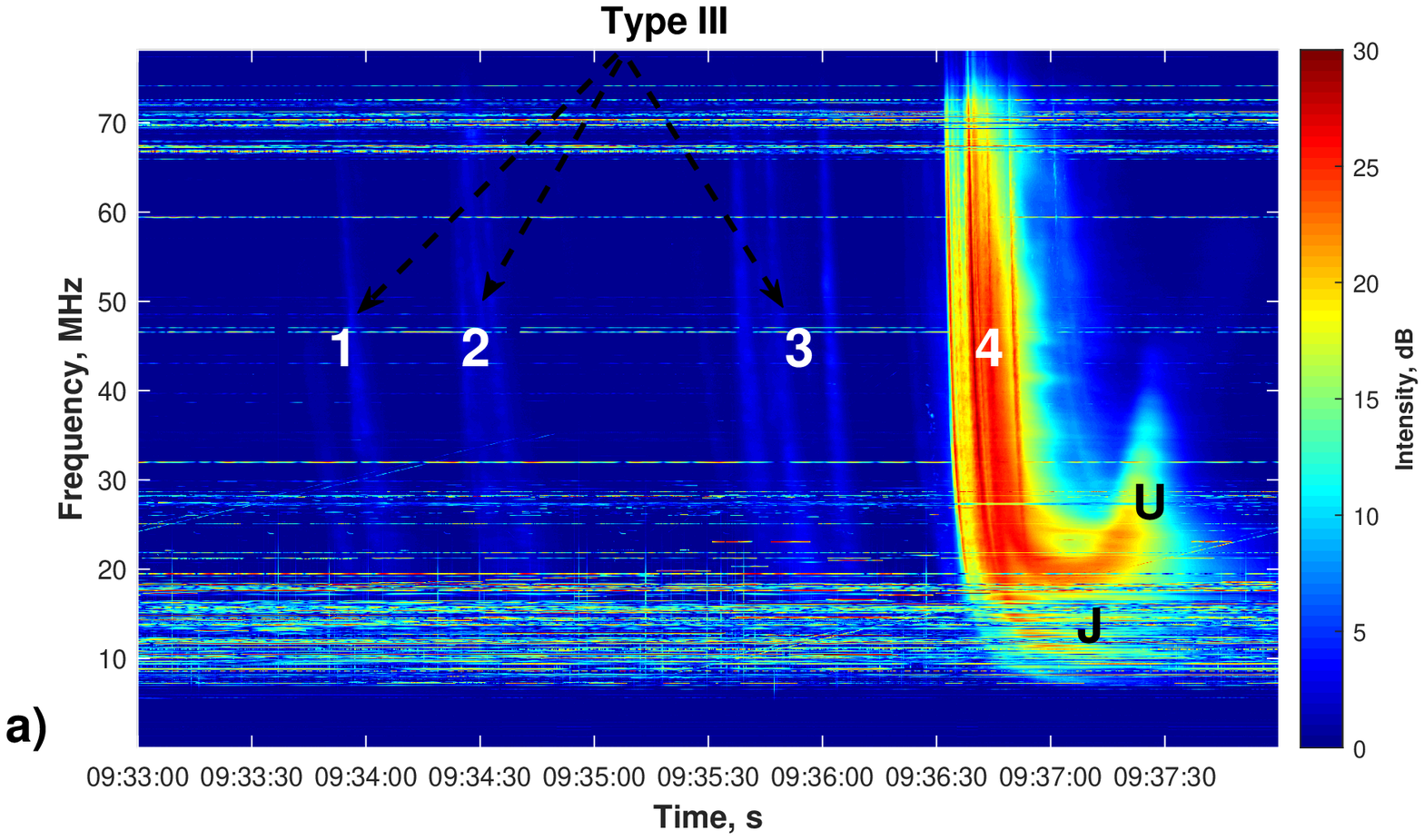}
         \includegraphics[width=14cm,keepaspectratio]{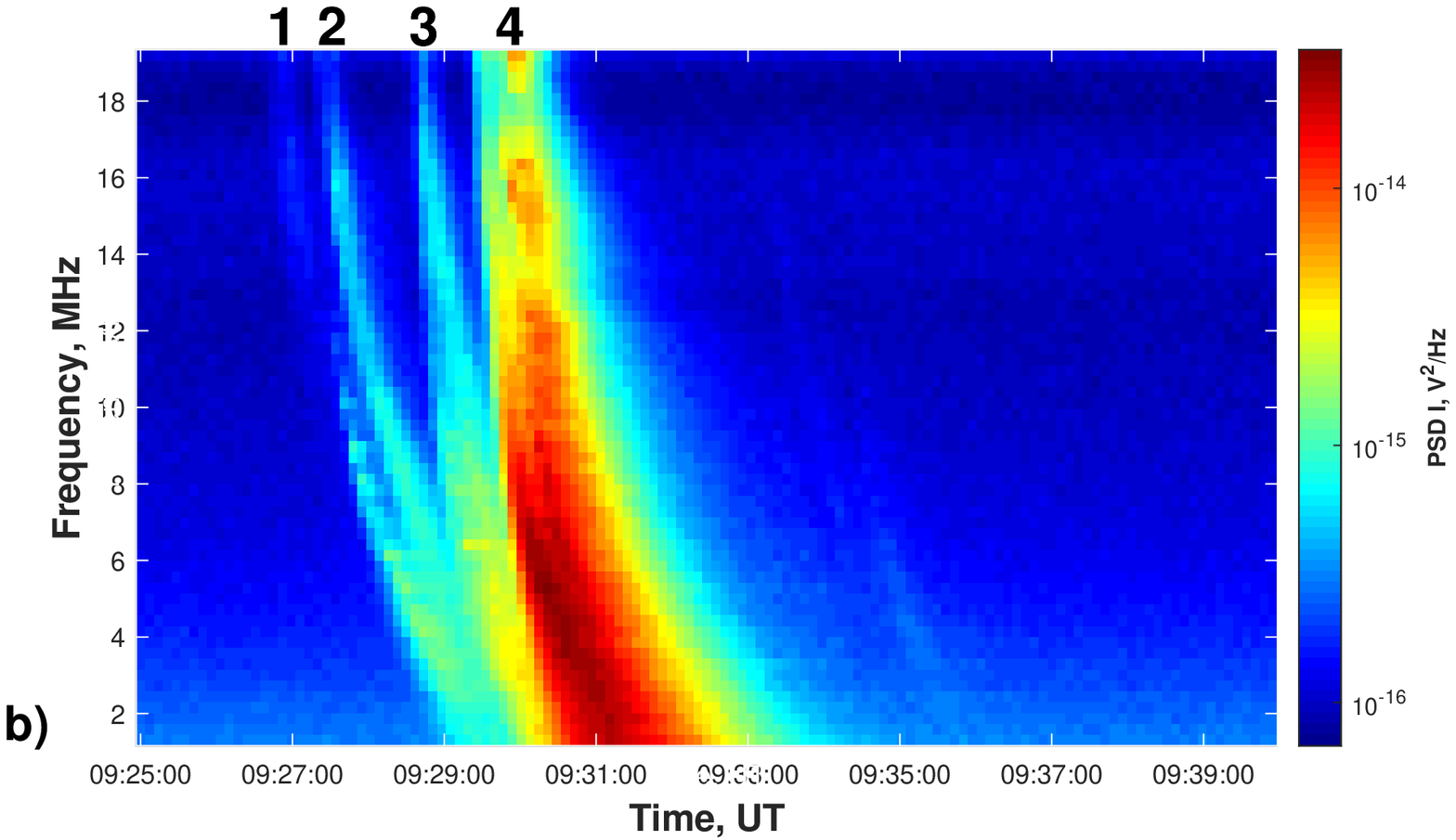}
\end{center}
\caption{Dynamic spectra of solar radio bursts on June 5, 2020 observed with GURT (a) and PSP (b). The letters III, U, and J denote the types of solar bursts, whereas the numbers indicate separate groups of type III bursts.}
\label{fig2} 
\end{figure*}

Often bipolar magnetic fields on the Sun are responsible for  type U and J solar radio bursts \citep{Fokker70,Stone71,Caroubalos73,Suzuki78,Leblanc83,Leblanc85,Aschwanden92,Yao97}. This type of radio burst was recorded on June 5, 2020. As it has been shown by \citet{Stanislavsky21}, this observation gave the first detection of the solar U+III association (Fig.~\ref{fig2}). The phenomenon was predicted in the paper by \citet{Reid17}, wherein they discuss how electron beams with different densities can generate radio emission in the form of only type III bursts as well as both type III and type U bursts. The latter regime is typical for high initial electron beam densities. In this case, the magnetic loop should be large enough for propagation effects, making the electron beam unstable, until it reaches the loop top and turns back to the Sun.

\section{Data analysis}
\subsection{Spectral measurements}
\begin{figure*}[!h]
\begin{center}
         \includegraphics[width=14cm,keepaspectratio]{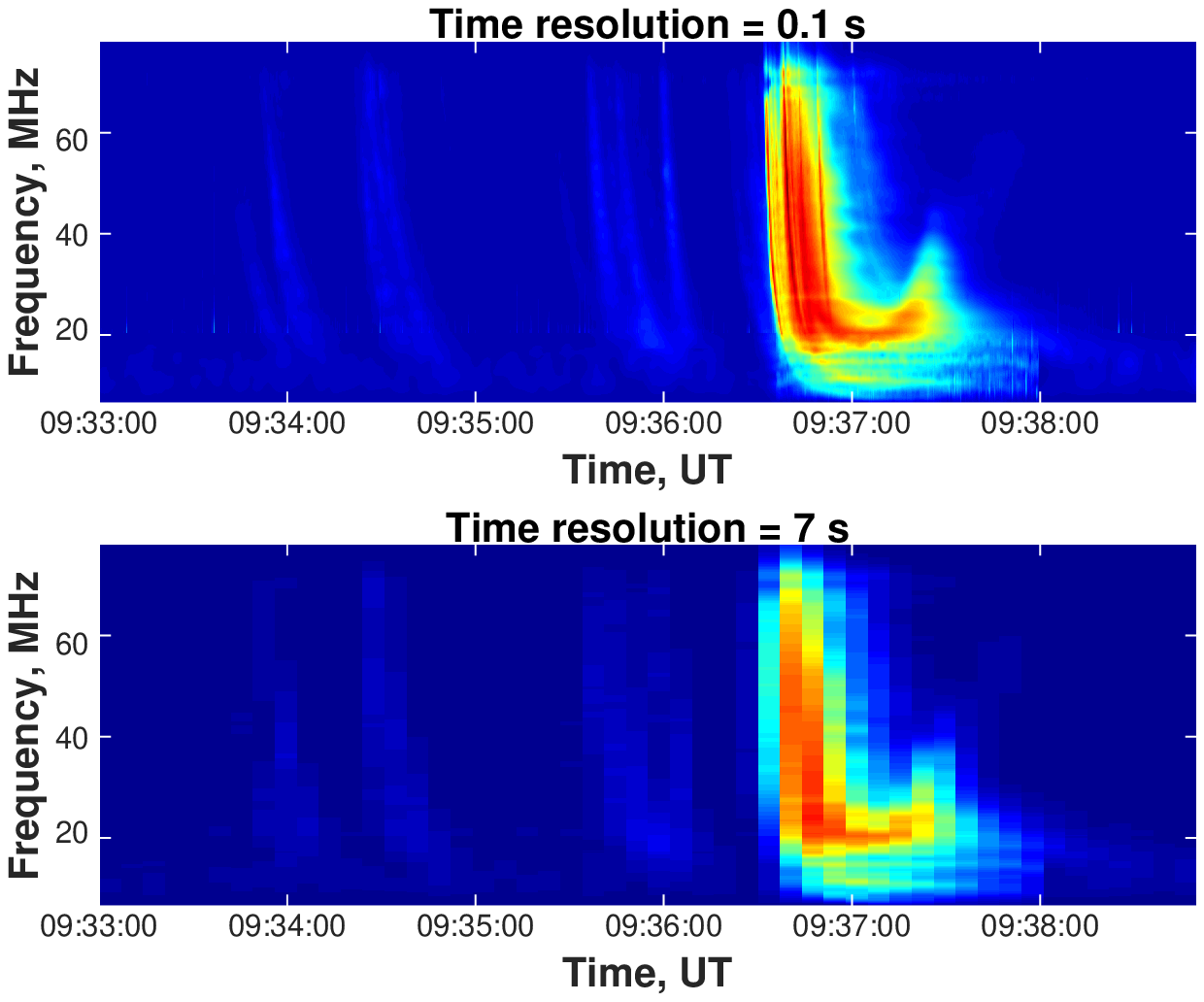}
\end{center}
\caption{Dynamic spectrum of solar bursts at 09:33:20 UT on June 5, 2020 after the RFI mitigation at 8--20 MHz so that the original case has  100 ms in time resolution whereas the averaged spectrum with 7 s resolution degrades.}
\label{fig2a} 
\end{figure*}

According to the GURT records (Fig.~\ref{fig2} a), the strong U+J+III bursts were surrounded by weaker type III bursts caused by electron beams with moderate densities. Although radio frequency interference (RFI) was too intensive and dense at 8--20 MHz for the GURT observations on June 5, 2020, the strong bursts can be isolated from RFI, using a special procedure of cleaning. The accurate and efficient RFI mitigation method was based on an Asymmetrically reweighted Penalized Least Squares (ArPLS) algorithm of the baseline estimation \citep{Baek15}. RFI forms sharp narrowband peaks, whereas solar bursts of U, J, and III types are broadband and prolonged. Using the algorithm, one can separate strong solar radio bursts from RFI. For weaker bursts, it is more difficult to do this as their spectrum degrades. Figure~\ref{fig2a} shows the result after such cleaning. Based on the procedure, we notice that the strong bursts at 09:36:30 UT not only indicate the association  of type U and III bursts, but also the J-type burst. Thus, the top of the U-type burst reached $\sim$18 MHz, the J-type fragment was at $\sim$15 MHz, and probably below, but it is difficult to note from the GURT records because of strong RFI.

As the time resolution of GURT records is 100 ms, we can study the solar event in more detail. In particular, the spectrogram manifests a fine structure of this event, marked with numbers, as well as denoted by the letters III, U, and J for types of solar bursts. Each number indicates a group of solar bursts. Both first and second groups contain two type III bursts. The third group has four such bursts. The forth group is the solar U+J+III association, including the U--burst together with several interplanetary type III bursts.  PSP also observed this event about 7 minutes earlier (Fig.~\ref{fig2} b). Its spectral features require clarifications. Unfortunately, the PSP HFR had a time resolution of $\sim$7~s. Moreover, each receiver channel of this instrument has 64 frequency channels, and thus the HFR resolution is about 280 kHz. Therefore, each group is not distinguishable and looks like separate type III radio bursts. Their durations increase with decreasing frequency, but their frequency drift rates decrease. Their close location on the spectrogram leads to the groups of radio bursts almost merging at the lowest frequencies of LFR, propped by a low--frequency cutoff. Although the frequency range and low time--frequency resolution of the PSP instrument does not permit one to consider the U+J+III association in detail, their fragment was confidently detected. The impact of lower time resolution on the GURT radio spectrum is presented in Fig.~\ref{fig2a}, where we have reduced it to 7 s, but kept the frequency resolution. Frequency drift rates of spectral patterns for each group (Fig.~\ref{fig2} a) are presented in Table~\ref{table1}. In general it can be argued that their values are typical for the type III radio bursts. Table~\ref{table2} describes frequency drift rates of four spectral patterns for the PSP HFR records. Although the tables use various frequency ranges and differ in frequency-time resolution, they are in agreement with each other. As it is well know \citep{Reid14}, the frequency drift rates of solar type III bursts at lower frequencies slow down. 

\begin{figure*}[!h]
\begin{center}
         \includegraphics[width=12cm,keepaspectratio]{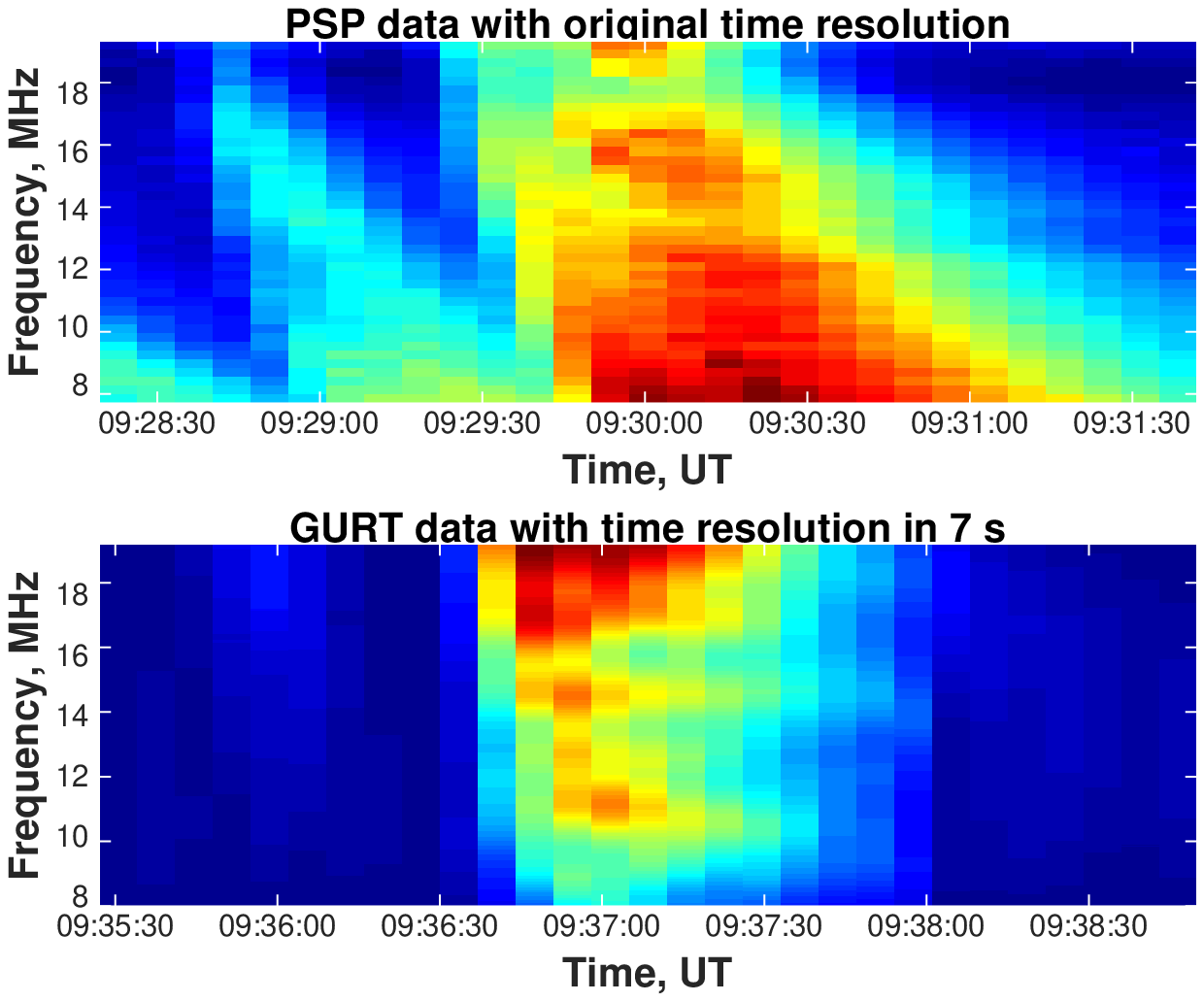}
\end{center}
\caption{Dynamic spectrum for a fragment of solar bursts on June 5, 2020 at 8--19 MHz, in which we compare the spectra from PSP and GURT records in time resolution.}
\label{fig2top} 
\end{figure*}

Next, Figure~\ref{fig2top} presents spectral fragments of the solar event (most power bursts) from the PSP and GURT records in the same frequency range (8--19 MHz) with the same time resolution. To see their similarities, several differences between the observations should be noticed, including the following: the frequency of 8 MHz being the boundary for the GURT telescope, while 19.2 MHz being that for the PSP radio instrument. In both limiting cases, the sensitivity and effectiveness for the receipt of radio signals turn out to be reduced. The following effects are seen on the fragments: the GURT array detects radio emission above 10 MHz in the best way; for PSP, the frequencies below 18 MHz are better for reception. However, the ground-based radio telescope was faced with a significant problem, caused by RFI at 8--20 MHz, which can only be partially solved by cleaning. One more point to keep in mind would be the different positions of the radio instruments in space. Readers should recall that PSP was about 6.7 times closer to the Sun than terrestrial radio telescopes, but the sensitivity of the GURT array is 5 times higher. Sources of the solar bursts have the directivity of radiation. Electron beams responsible for type III bursts move radially, whereas beams giving the U-type burst, travel along loop-like trajectories. In the latter case, the directivity diagram can turn in space. Therefore, there is no hope for an exact coincidence of the spectral fragments.

\begin{table*}
\caption{Frequency drift rate parameters of $\frac{df}{dt} = - K\,(f/30\,\,\mathrm{MHz})^\nu$ ($f$ in MHz and $df/dt$ in MHz s$^{-1}$) obtained for chosen patterns of the solar bursts shown in Fig.~\ref{fig2}a at 20 -- 80 MHz.}
\label{table1} 
\centering
\small
\begin{tabular}{c | c c | c c | c c | c c c c c}
\hline\hline
\multicolumn{1}{c|}{Groups} & \multicolumn{2}{|c|}{I} & \multicolumn{2}{|c|}{II} & \multicolumn{2}{|c|}{III} & \multicolumn{5}{|c}{IV} \\ 
\hline
Patterns & 1 & 2 & 3 & 4 & 5 & 6 & 7 & 8 & 9 & 10 & 11\\
\hline
$K$, MHz s$^{-1}$ & 2.38 & 1.89 & 1.92 & 2.45 & 3.39 & 5.25 & 6.32 & 6.4 & 4.31 & 6.92 & 7.65\\ 
$\nu$ & 1.18 & 2.08 & 2.63 & 1.68 & 1.39 & 1.31 & 2.1 & 1.9 & 1.73 & 1.75 & 1.51\\  
\hline 
\end{tabular}
\end{table*}

\begin{table*}
\caption{Frequency drift rate parameters of $\frac{df}{dt} = - K\,(f/30\,\,\mathrm{MHz})^\nu$ obtained for four patterns of the solar bursts shown in Fig.~\ref{fig2}b at 2 -- 19 MHz.}
\label{table2} 
\centering
\small
\begin{tabular}{c | c | c | c | c }
\hline\hline
\multicolumn{1}{c|}{Groups} & \multicolumn{1}{|c|}{I} & \multicolumn{1}{|c|}{II} & \multicolumn{1}{|c|}{III} & \multicolumn{1}{|c}{IV} \\ 
\hline
$K$, MHz s$^{-1}$ & 0.68 & 0.75 & 0.9 & 0.81\\ 
$\nu$ & 1.57 & 1.63 & 1.54 & 1.58 \\  
\hline 
\end{tabular}
\end{table*}

\subsection{Models of electron density in the solar corona}
The main feature of the solar U+J+III association is that the bursts started with $\sim$84 MHz. This allowed \citet{Stanislavsky21} to adjust the electron density model of solar corona above AR 12765. The point is that the corona of the quiet Sun can be described by the Newkirk ($\alpha$--fold) model \citep{Newkirk61}, in which the radial change of the electron density takes the following form:
\begin{equation}
n_e(r)=\alpha\cdot 4.2\cdot 10^4\cdot\,10^{(4.32/r)}\,,\label{eq1}
\end{equation}
where $n_e(r)$ is the electron density of coronal plasma in cm$^{-3}$, varying from the heliocentric height $r$ in solar radii. Here, $\alpha$ is the parameter dependent on the considered region on the Sun. If the region is quiet equatorial, then $\alpha=1$. In dense loops $\alpha$ = 4, and in extremely dense loops the parameter can reach 10 \citep{Koutchmy94}. Therefore, the model (\ref{eq1}) is called $\alpha$--fold \citep{Mann18}. The direct finding of $\alpha$ from radio observations is often difficult, but the solar U+J+III association makes it possible. The starting height \citep{Reid17}, with which the U--burst is nascent, is 0.6$R_\odot$ from the solar photosphere, and the burst begins with $\sim$85 MHz. The frequency was obtained from e--Callisto observations. We note that the RSTN observations give similar results, about 83.66 $\pm$ 0.26 MHz. This leads $\alpha\approx 4$ \citep{Stanislavsky21}. Then according to the cleaned spectrogram of Fig.~\ref{fig2a}, the top of the U--burst was at $\sim$18 MHz, that is the height reached about 3.1$R_\odot$. In the pivot point, the instantaneous bandwidth was equal to $\sim$5--7 MHz. This shows that the loop width at the top was about 0.53$R_\odot$. The results determine an approximate geometry of the position of solar radio burst sources  observed from different points around the Sun and received with STEREO--A, PSP, Wind/WAVES, and ground--based instruments. 

However, it should be pointed out that the Newkirk model is hydrostatic. At the distance of more than 3$R_\odot$, where the supersonic, superalfv\'enic solar wind sets up and the corona cannot be in hydrostatic equilibrium \citep{Alissandrakis20}, the model can overestimate the coronal electron density that makes the distances from the Sun too high in the analysis of low--frequency observations. In this context, the models of \cite{Leblanc98} and \cite{Mann99} are better. Particularly, the latter used a special solution from Parker's wind equation, which covers a range from the low corona up to 5 au. It is no coincidence that many authors multiplied the model's density by a factor of 2--4 or more, justified by the fact that in the burst environment, the coronal density is higher than in the case of the quiet Sun. Spherically and elliptically symmetrical representations on electron density of the solar corona have a number of restrictions. One of them concerns the observations of radio bursts initiated by ARs behind the limb for observers. Although under the conditions the radio emission cannot be detected by the observer, it is detected in real observations \citep{Stanislavsky17a,Stanislavsky17b}. Therefore, in this case the electron density model should by asymmetric. For simplicity, the factor is just introduced, though one can go about this in other ways. Readers should take note of how only the studies of \cite{Mann18} show that the factor $\alpha$ varied with frequency.

\begin{figure*}[!h]
\begin{center}
   \includegraphics[width=14cm,keepaspectratio]{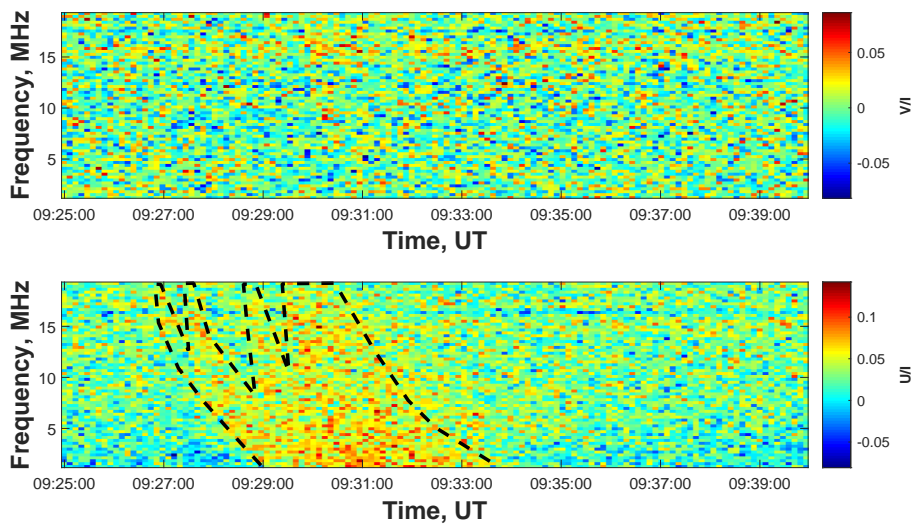}
\end{center}
\caption{Solar bursts with linear polarization at 09:27:30 UT on June 5, 2020. The dotted line shows boundaries of their Stokes intensity $I$ for comparison.}
\label{fig2b} 
\end{figure*}

\subsection{Polarization measurements}
The PSP radio data contain both auto and cross spectra. This allowed us to find the Stokes $I$ (Stokes intensity), $Q$, $U$, and $V$ parameters, representing the linear ($Q$, $U$) and circular ($V$) polarization for received radio emission. Following \cite{Pulupa20}, their values read as follows:
\newcommand*{\autoa}{{V}_{12} {V}_{12}^*}
\newcommand*{\autob}{{V}_{34} {V}_{34}^*}
\newcommand*{\xspec}{{V}_{12} {V}_{34}^*}

\begin{equation}
I = \autoa + \autob
\end{equation}
\begin{equation}
Q = \autoa - \autob
\end{equation}
\begin{equation}
U = \xspec + (\xspec)^*
\end{equation}
\begin{equation}
iV = \xspec - (\xspec)^*
,\end{equation}
where $\autoa$ and $\autob$ are the auto spectra from the $V1-V2$ and $V3-V4$ channels, respectively, and $\xspec$ is the cross spectrum. Figure~\ref{fig2b} shows $V/I$ and $U/I$  spectrograms for the solar event with clear signatures of linear polarization rather than a circular one. Readers should recall that the observations of \cite{Pulupa20} and their analysis display circular polarization for the type III bursts. It is clear why; the presence of a magnetic field in a solar plasma wipes out linear polarization \citep{Grog73,Boischot75,Alissandrakis21}. 

In our study, we have probably detected another case. The preservation of linear polarization in the radio emission of solar bursts can be explained by the fact that on June 5, 2020, PSP observed the radio source sideways, and the magnetic field $B$ in the solar corona is directed radially. When looking perpendicular to $B$, the natural radiation modes could be predominantly linear. However, it cannot be ruled out that for PSP radio measurements, the linear polarization Stokes parameters are sensitive to small differences in the electrical effective lengths of the PSP dipole antennas, and/or nonideal antenna geometry, namely two dipole antennas are not perfectly orthogonal due to the accommodation of other instruments on the spacecraft (private communication from M. Pulupa). The Stokes parameter $V$ is a different case \citep{Lecaceux11}: some properties of short dipole antennas allow one to make reasonable estimates for $V$ without fully accounting for all of these effects. But this begs the question of why circular polarization of this solar event was not noticeable. 

As for GURT, the instrument permits polarization measurements, but this feature has not been fully implemented in the observations. Nevertheless, we have noticed qualitative effects such as ionospheric fringes in our cross spectra and circular polarization of the solar bursts. As the measurements were not calibrated right until the end, we do not give them  here in detail.

\begin{figure*}[!h]
\begin{center}
        \includegraphics[width=14cm,keepaspectratio]{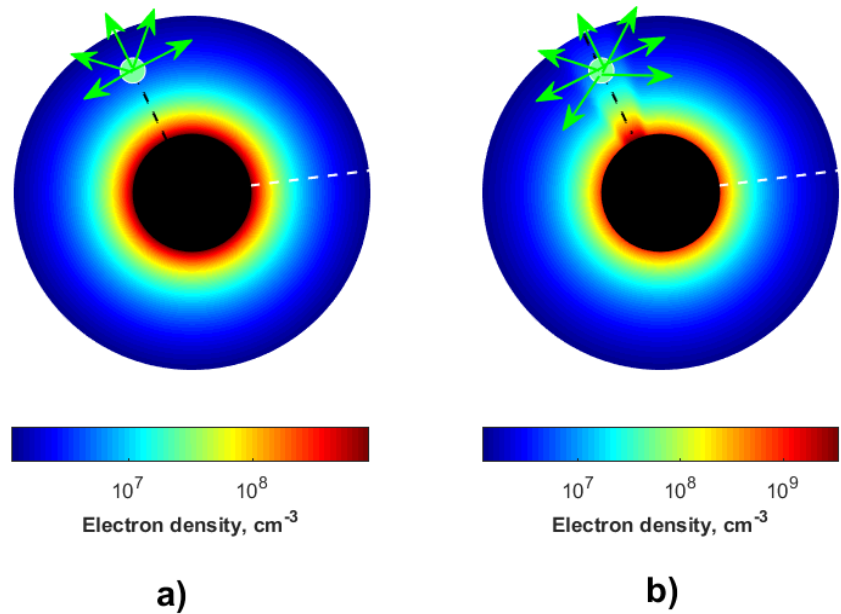}
\end{center}
\caption{Sketch representing the radio emitter position (green filled circle on the black dotted line) and the direction to PSP (white dotted line) with respect to the Sun: a) spherically symmetric corona and b) corona with a dense sector caused by AR. The green arrows show possible ray path propagation of fundamental emission. The black filled circle shows the solar photosphere.}
\label{fig3} 
\end{figure*}

\section{Results}
The path of radio emission rays in the solar corona satisfies Snell's law so that variations of the refractive index with height can change the direction of propagation. If the plasma density monotonically decreases in the solar corona with distance from the photosphere, and if the case is spherically symmetric, then the radiation caused by behind--the--limb burst--active ARs cannot be received with the observer's side because of radio emission reflection in the direction out of the observer \citep{Bracewell56}. In this case, the ray path between the radio source and any terrestrial observer meets a more dense region and therefore deviates away from the observer (Fig.~\ref{fig3} a). However, the solar corona is heterogeneous and nonstationary in nature. Due to solar activity, sunspots appear, develop, and disappear; also, flares and coronal mass ejections occur similar to other phenomena in the solar photosphere and corona. Consequently, the corona turns out to be changeable in density, not only with height, but also at spherical angles. Therefore, if tunnel--like cavities with low density emerge in the solar corona, the radiation will be refracted in the formations rather than reflected. In this case, the radio emission, initiated by behind--the-limb ARs, can be detected by observers \citep{Stanislavsky17a}. The AR 12765, which activated electron beams responsible for the solar bursts on June 5, 20020, was behind the solar limb for PSP instruments. Their detection can be explained in a simple way. The solar side to PSP was without sunspots. Thus, the corona density was low (quiet region). At the same time, the radio sources of solar bursts were located in a dense loop. This means that the radio waves spread from denser plasma to less dense plasma (Fig.~\ref{fig3} b). Such conditions in the solar corona state were favorable for radio observations of activity from the behind-limb AR for PSP. One of the main features of solar radio events from behind--the--limb ARs is their high--frequency cutoff \citep{Stanislavsky17a}. The cause of this effect is that radiating sources are occulted by the solar corona for observers, and radio emission with high frequencies originates closer to the photosphere. As applied to the solar bursts on 5 June 5, 2020, the cutoff was not noticed because the highest frequency of radio observations with PSP is 19.2 MHz. It looks as though the frequency range of PSP was not enough to establish the effect. As an example, we mentioned solar radio records on August 18, 2012, which show the high--frequency cutoff near 26 MHz at 08:05 UT and about 40 MHz at 13:03 UT \citep{Stanislavsky17b}.

The observations of the solar event on June 5, 2020 at $\sim$09:30 UT have been made by both space-- and ground--based radio instruments. Their different locations in outer space around the Sun allowed us to estimate the angular width of the directivity of solar type III, J, and U bursts from 6 MHz to 19 MHz. The frequency range is common for space-- and ground--based observations. Using a special antenna intended for the sensitive receipt of radio emission at 4--40 MHz \citep{Stanislavsky21}, the solar event was detected in extremely low frequencies close to one ionospheric cutoff (about 6 MHz on that day). Assuming that the directivity diagram is symmetric with respect to the radial motion direction of electron beams from AR 12765,  the angular width, as applied to the directivity of the solar radio bursts, was more than 212$^\circ$. This result is consistent with  previous investigations for the directivity diagram of solar type III bursts \citep[see, for example,][]{Bonnin08, Konov13}. It is very likely that if the satellite STEREO-B was workable at that time, then it could receive radio emission from this solar event too.

\section{Conclusions}
In this paper we have shown the mutual usefulness of joint observations made with the help of ground--based radio telescopes and the PSP radio instrument. On the one side, the GURT, NDA, RSTN, and e--Callisto observations tracked down the events from high frequencies to the frequency of the ionosphere cutoff. Their high time--frequency resolution and sensitivity permitted us to detect the U+J+III association surrounded by weaker type III solar bursts. Using the generation mechanism of type U+J+III bursts, we refined the density model of the solar corona, which was used for the interpretation of observations obtained with PSP located in perihelion behind the Sun for observers on Earth. Knowing the PSP spacecraft position near the Sun and delays of radio waves between space-- and ground--based records, we found the corresponding radio responses on the PSP spectrogram. On the other hand, on June 5, 2020, the Sun had minimal solar activity and only woke up for a new cycle. The small number of sunspots created favorable conditions (quiet regions with low densities) in the solar corona for observations of solar bursts (emitted from denser regions) at low frequencies despite behind--the--limb locations of ARs. Although lower time--frequency resolution (in comparison with ground--based instruments) of the PSP receiver did not allow us to select solar type III bursts in each group, the groups were detected separately. Moreover, the analysis demonstrates that the type III bursts were interplanetary due to the radio observations with the spacecrafts STEREO--A and Wind/WAVES.\\


\noindent{\it Acknowledgements.} AAS thanks the Polish National Agency for Academic Exchange for their support (NAWA PPN/ULM/2019/1/00087/DEC/1). AAK acknowledges support from the project RVO:67985815 and Grant: 20--09922J of the Czech Science Foundation. INB and SNY are thankful for the support of Grant 0121U109536 from the National Academy of Sciences of Ukraine. The authors thank the Wind/WAVES, STEREO, SOHO, SDO, RSTN, e--Callisto and NOAA teams for their instrument maintenance and open data access. Personally, we are also grateful to M. Pulupa for his useful remarks. The FIELDS experiment on the Parker Solar Probe spacecraft was designed and developed under NASA contract NNN06AA01C. PSP/FIELDS data are publicly available at http://fields.ssl.berkeley.edu/data/. The authors thank the radio astronomy station of Nan\c{c}ay / Scientific Unit of Nan\c{c}ay of the Paris Observatory (USR 704-CNRS, supported by the University of Orleans, the OSUC and the Center Region in France) for providing access to NDA observations accessible online at https://www.obs-nancay.fr.






\begin{thebibliography}{0}
\expandafter\ifx\csname natexlab\endcsname\relax\def\natexlab#1{#1}\fi

\bibitem[{{Alissandrakis}}(2020)]{Alissandrakis20}
Alissandrakis, C. E. 2020, Front. Astron. Space Sci., 7, 574460

\bibitem[{{Alissandrakis \& Gary}}(2021)]{Alissandrakis21}
Alissandrakis, C. E. \& Gary, D. E. 2021, Front. Astron. Space Sci., 7, 591075

\bibitem[{{Aschwanden} {et~al.}}(1992)]{Aschwanden92}
{Aschwanden}, M. J., {Bastian}, T. S., {Benz}, A. O., \& {Brosius}, J. W. 1992, \apj, 391, 380

\bibitem[{{Baek} {et~al.}}(2015)]{Baek15}
{Baek}, S.-J., {Park}, A., {Ahn}, Y.-J., \& {Choo}, J. 2015, Analyst, 140, 250

\bibitem[{{Bale} {et~al.}}(2016)]{Bale16}
{Bale}, S. D., {Goetz}, K., {Harvey}, P. R., et~al. 2016, \ssr, 204, 49

\bibitem[{{Boischot \& Lecacheux}}(1975)]{Boischot75}
{Boischot}, A., \& {Lecacheux}, A. 1975, \aap, 40, 55

\bibitem[{{Bonnin} {et~al.}}(2008)]{Bonnin08}
{Bonnin}, X., {Hoang}, S., \& {Maksimovic}, M. 2008, \aap, 489, 419

\bibitem[{{Bracewell \& Preston}}(1956)]{Bracewell56}
{Bracewell}, R. N., \& {Preston}, G. W. 1956, \apj, 123, 14

\bibitem[{{Caroubalos} {et~al.}}(1973)]{Caroubalos73}
{Caroubalos}, C., {Couturier}, P., \& {Prokakis}, T. 1973, \aap, 23, 131

\bibitem[{{Fokker}}(1970)]{Fokker70}
{Fokker}, A. D. 1970, \solphys, 11, 92

\bibitem[{{Fox} {et~al.}}(2016)]{Fox16}
{Fox}, N. J., {Velli}, M. C., {Bale}, S. D., et~al. 2016, \ssr, 204, 7

\bibitem[{{Grognard \& McLean}}(1973)]{Grog73}
{Grognard}, R. J. M., \& {McLean}, D. J. 1973, \solphys, 29, 149

\bibitem[{{Konovalenko} {et~al.}}(2013)]{Konov13}
{Konovalenko}, A. A., {Stanislavsky}, A. A., {Rucker}, H. O., et~al. 2013, Experim. Astron., 36, 137

\bibitem[{{Konovalenko} {et~al.}}(2016)]{Konov16}
{Konovalenko}, A., {Sodin}, L., {Zakharenko}, V., et~al. 2016, Experim. Astron., 42, 11

\bibitem[{{Koutchmy}}(1994)]{Koutchmy94}
{Koutchmy}, S. 1994, Adv. Space Res., 14, 29

\bibitem[{{Krupar} {et~al.}}(2020)]{Krupar20}
{Krupar}, V., {Szabo}, A., {Maksimovic}, M., et~al. 2020, \apjs, 246, 57

\bibitem[{{Leblanc} {et~al.}}(1983)]{Leblanc83}
{Leblanc}, Y., {Poquerusse}, M., {Aubier}, M. G. 1983, \aap, 123, 307

\bibitem[{{Leblanc \& Hoyos}}(1985)]{Leblanc85}
{Leblanc}, Y., \& {Hoyos}, M. 1985, \aap, 143, 365

\bibitem[{{Leblanc} {et~al.}}(1998)]{Leblanc98}
{Leblanc}, Y., {Dulk}, G. A., \& {Bougeret}, J.-L. 1998, \solphys, 183, 165

\bibitem[{{Lecacheux}}(2011)]{Lecaceux11}
{Lecacheux}, A. 2011, In: Rucker, H. O., Kurth, W. S., Louarn, P., \& Fischer G. (eds.) {\it Proceedings of the 7th International Workshop on Planetary, Solar and Heliospheric Radio Emissions (PRE VII)}, Austrian Academy Sciences Press, Vienna, 13

\bibitem[{{Ma} {et~al.}}(2021)]{Ma21}
{Ma}, B., {Chen}, L., {Wu}, D., \& {Bale}, S.D. 2021, \apjl, 913, L1

\bibitem[{{Mann} {et~al.}}(1999)]{Mann99}
{Mann}, G., {Jansen}, F., {MacDowall}, R. J., {Kaiser}, M. L., \& {Stone}, R. G. C. 1999, \aap, 348, 614

\bibitem[{{Mann} {et~al.}}(2018)]{Mann18}
{Mann}, G., {Breitling}, F., {Vocks}, C., et~al. 2018, \aap, 611, A57

\bibitem[{{Melnik} {et~al.}}(2021)]{Melnik21}
{Melnik}, V. N., {Brazhenko}, A. I., {Konovalenko}, A. A., {Frantsuzenko}, A. V., {Yerin}, S. M., {Dorovskyy}, V. V., \& {Bubnov}, I. M. 2021, \solphys, 296, 9

\bibitem[{{Newkirk}}(1961)]{Newkirk61}
{Newkirk}, G. 1961, \apj 133, 983

\bibitem[{{Pulupa} {et~al.}}(2017)]{Pulupa17}
{Pulupa}, M., {Bale}, S. D., {Bonnell}, J. W., et~al. 2017, JGR Space Phys., 122, 2836

\bibitem[{{Pulupa} {et~al.}}(2020)]{Pulupa20}
{Pulupa}, M., {Bale}, S. D., Badman, S. T., et~al. 2020, \apjs, 246, 49

\bibitem[{{Reid \& Kontar}}(2017)]{Reid17} 
{Reid}, H. A. S., \& {Kontar}, E. P. 2017, \aap, 606, A141

\bibitem[{{Reid \& Ratcliffe}}(2014)]{Reid14}
{Reid}, H. A. S., \& {Ratcliffe}, H. 2014, Res. Astron. Astrophys., 14, 773

\bibitem[{{Stanislavsky}}(2017a)]{Stanislavsky17a}
{Stanislavsky}, A.A. 2017a, Astronom. Nachr., 338, 407

\bibitem[{{Stanislavsky} {et~al.}}(2017b)]{Stanislavsky17b}
{Stanislavsky}, A. A., {Konovalenko}, A .A., {Volvach}, Ya. S., \& {Koval}, A.A. 2017b, In: Fischer, G., Mann., G., \& Zarka, Ph. (eds.) {\it Proceedings of the 8th International Workshop on Planetary, Solar and Heliospheric Radio Emissions (PRE VIII)}, Austrian Academy of Sciences Press, Vienna, 391

\bibitem[{{Stanislavsky} {et~al.}}(2021)]{Stanislavsky21}
{Stanislavsky}, L. A., {Bubnov}, I. N., {Konovalenko}, A. A., {Tokarsky}, P. L., \& {Yerin}, S. N. 2021, Res. Astron. Astrophys., 21, 187

\bibitem[{{Stone \& Fainberg}}(1971)]{Stone71}
{Stone}, R. G., \& {Fainberg}, J. 1971, \solphys, 20, 106

\bibitem[{{Suzuki}}(1978)]{Suzuki78}
{Suzuki}, S. 1978, \solphys, 57, 415

\bibitem[{{Wild}}(1950)]{wild50}
{Wild}, J. P. 1950, Austr. J. Sci. Res., A3, 541

\bibitem[{{Yao} {et~al.}}(1997)]{Yao97}
{Yao}, J.-X., {Yu}, X.-F., {Tlamicha}, A., \& {Wei}, F.-S. 1997, Adv. Space Res., 20, 2351

\bibitem[{{Zakharenko} {et~al.}}(2016)]{Zakharenko16}
{Zakharenko}, V., {Konovalenko}, A., {Zarka}, P., et~al. 2016,  Journal of Astronomical Instrumentation, 5, 1641010

\bibitem[{{Zheleznyakov}}(1970)]{zhel70}
{Zheleznyakov}, V. V. 1970, {\it Radio Emission of the Sun and Planets}, (Oxford: Pergamon Press) 

\end{thebibliography}
\end{document}